\documentclass[conference]{IEEEtran}
\IEEEoverridecommandlockouts

\usepackage{graphicx}
\usepackage[utf8]{inputenc}
\usepackage[T1]{fontenc}
\usepackage{listings}
	\lstdefinelanguage{diff}{
    basicstyle=\ttfamily\small,
    morecomment=[f][\color{diffstart}]{@@},
    morecomment=[f][\color{diffincl}]{+\ },
    morecomment=[f][\color{diffrem}]{-\ },
  }
\usepackage[svgnames]{xcolor}
  \definecolor{diffstart}{named}{Grey}
  \definecolor{diffincl}{named}{Green}
  \definecolor{diffrem}{named}{OrangeRed}
\usepackage{color}
\usepackage{xspace}
\usepackage[inline]{enumitem}
\usepackage{graphicx}
\usepackage{tcolorbox}
\usepackage{hyperref}
\usepackage{multirow}
\usepackage[sorting=none,doi=false,isbn=false,url=false]{biblatex}
\usepackage{caption} 
\newtheorem{Finding}{\textbf{Finding}}
\newtheorem{Implication}{\textbf{Implication}}

\newcommand{\ie}{\textit{i.e.,}\xspace}
\newcommand{\eg}{\textit{e.g.,}\xspace}
\newcommand{\etc}{\textit{etc.}\xspace}
\newcommand{\etal}{\textit{et al.}\xspace}

\bibliography{references.bib}
\graphicspath{ {images/} }

\begin{document}

\title{The Remarkable Role of Similarity in Redundancy-based Program Repair
}

\author{
\IEEEauthorblockN{Zimin Chen}
\IEEEauthorblockA{\textit{KTH Royal Institute of Technology}\\
zimin@kth.se}
\and
\IEEEauthorblockN{Martin Monperrus}
\IEEEauthorblockA{\textit{KTH Royal Institute of Technology}\\
martin.monperrus@csc.kth.se}
}

\maketitle

\begin{abstract}
Recently, there have been original attempts to use the concept of ``code similarity'' in program repair, suggesting that similarity analysis has an important role in the repair process. However, there is no dedicated work to characterize and quantify the role of similarity in redundancy-based program repair, where the patch is composed from source code taken from somewhere else. This is where our paper makes a major contribution: we perform a deep and systematic analysis of the role of code similarity during the exploration of the repair search space. We define and set up a large-scale experiment based on four code similarity metrics that capture different similarities: character, token, semantic and structure similarity.
Overall, we have computed 56 million similarity score over 15 million source code components. We show that with similarity analysis, at least 90\% of search space can be ignored to find the correct patch. Code similarity is capable of ranking the correct repair ingredient first in 4 - 33 \% of the considered cases. 
\end{abstract}

\section{Introduction}

In program repair research, one can identify at least two broad categories of papers.
On the one hand, there are papers which propose new repair techniques based on a powerful innovative idea (such as \cite{capgen-icse18,hua2018towards,mechtaev2018semantic,van2018static}, to only mention recent ones).
On the other hand, there are papers which pursue the endeavor of understanding the core structure and challenges of the repair search space  (such as \cite{Martinez2013,qi2014strength,long2016analysis,le2018overfitting}). The work presented here fits in this latter category.

Redundancy-based program repair consists of repairing programs with code taken from elsewhere, for instance from the application under repair.
Redundancy-based program repair is fundamental to the repair community,
early approaches such as GenProg \cite{weimer2009automatically} and major state-of-the-art approaches (e.g. \cite{capgen-icse18,SimFix-issta2018,hua2018towards}) both relied on redundancy.
While the empirical presence of redundancy has been studied \cite{Martinez:2014:FIA:2591062.2591114,barr2014plastic}, the search space of redundancy-based repair is little known. 

Recently, there have been original attempts to use the concept of similarity in redundancy-based repair techniques.
In ssFix, CapGen and SimFixx, different kinds of similarity analysis are part of the overall repair technique \cite{ssfix-ase17,capgen-icse18,SimFix-issta2018}.
ssFix uses TFIDF to compute the similarity between the buggy statement and its context \cite{ssfix-ase17}. 
CapGen prioritizes potential repair code snippets that have a similar context, with three similarities stacked in a clever manner \cite{capgen-icse18}.
SimFix heavily relies on name similarity (variable name and method name) to find code snippets that are present in other similar methods \cite{SimFix-issta2018}.
Given that those works report impressive results in terms of repair effectiveness, it is clear that similarity has an important role in the repair process.
However, there is no dedicated work to characterize and quantify the role of similarity in redundancy-based program repair. This is where our paper makes a major contribution: we perform a deep and systematic analysis of similarity analysis during the exploration of the search space.

Our study of similarity in redundancy-based program repair uses a strictly principled methodology.
First, we isolate the similarity component in a generic redundancy-based repair process. This isolation means that we simplify the repair algorithm as much as possible (no randomness, no fault localization and patch validation) so that similarity analysis becomes the main component of the repair process. Thus, we ensure that the observed effects on the search space are those of similarity analysis.
Second, we systematically qualify the similarity relationships that we consider: in the context of a replacement patch where one statement is replaced by another statement, we measure the similarity between the removed and the inserted code. We also consider the context around the removed and inserted statement and measure the similarity between statement contexts.
We systematically define four similarity metrics that capture either syntactic, semantic or tree similarity are used and investigated. To our knowledge, this is the first time that four different similarity metrics are thoroughly evaluated in the context of program repair. 

Then, we design and perform a unique series of  experiments.
The main goal of the experiments is to study how good  similarity analysis is at finding the correct repair ingredient in the search space. For evaluate similarity analysis, we create repair tasks from real commits, where each repair task is a ranking problem: given the buggy source code, return a list of repair ingredients ordered by the likelihood that they will fix the bug. We evaluate the importance of considering the context around the buggy statement for finding the correct repair ingredient. 

Our experimental results are clear-cut.
First, using similarity analysis is effective in reducing the search space: our experiments show that at least 90\% of the search space can be ignored when finding the correct patch. Considering the similarity metric TFIDF, the best search space reduction is 99.56\% averaging over all projects. This shows that the ranking of repair ingredients provided by similarity analysis is very effective.

Second, we show that including the context in similarity analysis makes the repair ingredient search even more powerful. When combining context-aware and context-less similarity analysis, the correct repair ingredient can be ranked higher, in 17\% of the cases, the correct repair ingredient is ranked first. 
Overall, our novel methodology enables us to have full control over the search and gives the community a unique,  deep understanding of the search space of redundancy-based program repair.

To sum up, we make the following contributions:
\begin{itemize}[leftmargin=*]
\item A principled conceptual framework to study the usage of code similarity analysis in redundancy-based program repair. 

\item Fundamental empirical findings of the remarkable performance of similarity in redundancy-based repair based on 56 million similarity comparisons. Using similarity analysis enables us to cut the search space between $91.87\%$ and $99.56\%$.

\item A systematic study of the importance of the repair context in redundancy-based program repair. Our results over 214 repair tasks show that taking into account the method body around the buggy statement as context enables to improve the effectiveness of similarity analysis. It cut the search space by an additional $24\%$ compared to without considering the context.

\item A statistical analysis of the rank distributions for the four considered similarity metrics. This confirms the correlation between purely syntactic based similarity analysis and suggests future work on combining different approaches together.

\end{itemize}

\section{Background}

\subsection{Terminology}
\label{sec:terminology}
The following concepts are used throughout this study:
\begin{itemize}
\item \textbf{Modification point:} It is a source code component that is suspicious of causing the bug \cite{martinez2019astor}. In this study, the modification point that we consider is a statement.
\item \textbf{Repair ingredient:} It is a source code component that can be used to fix the bug at the modification point. \eg by replacing the modification point with one repair ingredient. The repair ingredient that correctly fixes the bug is called the ``correct repair ingredient" \cite{white2017sorting}.
\item \textbf{Recipient context:} It is the context around the modification point  \cite{barr2014plastic}. In this study, the recipient context is a method enclosing the modification point.
\item \textbf{Donor context:} It is a source code snippet around a specific repair ingredient \cite{barr2014plastic}. In this study, 
we consider a method as donor context.
\end{itemize}

\subsection{Overview of Redundancy-based Program Repair}

Redundancy-based program repair consists of repairing programs with code taken from elsewhere, for instance from the application under repair. The insight behind it is that source code is very redundant \cite{gabel2010study}, which means that the same feature is implemented at multiple locations in slightly different ways. Consequently, a bug that affects a code snippet can be fixed by another code snippet.

Redundancy-based program repair is fundamental to the repair community as it originates from the inception of the field:  back in 2009, GenProg already leveraged redundancy \cite{weimer2009automatically}.
Today, major state-of-the-art approaches still heavily depend on redundancy (e.g. \cite{capgen-icse18,SimFix-issta2018,hua2018towards}).

The redundancy assumption --- the assumption that code may evolve from existing code that comes from somewhere else --- has been empirically verified. Martinez \etal and Barr \etal both verified its validity by analyzing thousands of past commits \cite{Martinez:2014:FIA:2591062.2591114, barr2014plastic}: between 3\% and 17\% of commits are only composed of existing code.
Recently, it has been reported that a redundancy-based approach successfully scales to an industrial project \cite{naitou2018toward}. 

Furthermore, Martinez \etal have refined this idea with the concept of ``redundancy scope'' \cite{Martinez:2014:FIA:2591062.2591114} , which defines the boundaries from which the repair ingredients are taken.
For instance, the file scope means that repair ingredients are only taken from the same file from which the suspicious faulty statement comes (this is the default scope of GenProg). For another example, the application scope means that the ingredients come from the whole application. In this paper, we focus on the application scope.

\subsection{Search Space of Redundancy-based Repair}
In theory, the search space of program repair is virtually infinite, composing of all possible edits on a program. 
Redundancy-based program repair is one way to dramatically reduce the search space.
In essence, it reduces the search space to the number of repair ingredients that already exist, where a repair ingredient is either a token, an AST node, a line or a full snippet.

While the search space under the redundancy assumption is effectively reduced, it may still be too large in practice. In essence, it depends on the redundancy scope. For instance, at the file scope, the search space size is proportional to the file size (usually hundreds of lines, at most thousands).
At the application scope, the size of the search space, with some simplification, is the number of lines in the application. Thus, for a 1 million LOC application, the search space is composed of 1 million elements (or a certain number proportional to it), which is quite large.
In practice, exploring the search space means picking an ingredient, compiling it, and executing test cases. Since there myriads of possible ingredients, this is too long: the search space of redundancy-based repair is often too big to be exhaustively explored.

Large search space introduces another problem, overfitting repair ingredients.
Overfitting repair ingredients are those who yield overfitting patches that pass all tests but fail to generalize to other buggy inputs \cite{le2018overfitting}. Long and Rinard found that overfitting repair ingredients are far more abundant than correct repair ingredients \cite{long2016analysis}. For instance, Qi \etal found that most patches generated by GenProg, RSRepair and AE, all being redundancy-based repair approaches, are overfitting patches \cite{qi2015analysis}. 

When it is too slow to exhaustively explore the search space of redundancy-based repair, what is the solution to only explore the relevant parts of the search space? This is the open question we systematically explore in this paper.

\subsection{Similarity Analysis for Redundancy-based Repair}

Major recent work in program repair has shown that similarity analysis is a useful concept for program repair, being present in ssFix \cite{ssfix-ase17}, CapGen \cite{capgen-icse18} and SimFix  \cite{SimFix-issta2018}. It is an assumption that source code that are similar to the buggy source code are more likely to contain the bug fix. For instance, let us consider the human patch of bug Math-75 from Defects4J \cite{Just:2014:DDE:2610384.2628055} in \autoref{humanpatchmath75}.
We can see that the inserted statement and the removed statement are syntactically very similar. However, ssFix, CapGen and Simfix, they all consider different forms of similarity and use different metrics for similarity analysis.

For instance, ssFix \cite{ssfix-ase17} finds suspicious statements that are more likely to be faulty with fault localization. Then, for each suspicious statement, the context and the suspicious statement is extracted to identify similar code chunk in the codebase with Lucene's default TFIDF model. The ingredient is then obtained from the similar code chunk to generate a patch. \cite{xin2019revisiting} have revisited ssFix and created sharpFix, where the context and the similarity metric are changed. They did obtain better result, but the idea of using similarity analysis is very much the same.

On the other hand, CapGen \cite{capgen-icse18} considers context similarity, name similarity and dependency similarity between the suspicious code fragments and the ingredients at the expression level. They extend the similarity analysis with results from their empirical study which calculates the frequency between different mutation operators (replacement, insertion and deletion) and the type of involved code fragment (method invocation, if statement and \etc) to prioritize ingredients that are more likely to fix the bug.

SimFix \cite{SimFix-issta2018} does also consider structure similarity, variable name similarity and method name similarity between the suspicious code snippet and donor snippet (not a method in this case). The top 100 most similar donor snippet is selected and repair ingredients are extracted for the repair process.

We can also see that these program repair approaches consider the context around the repair ingredients. In those works, the similarity metric is calculated between the context around the modification point (the recipient context) and the context around the repair ingredients (the donor context). For example, SimFix considers a constant number of statements before and after the modification point as the context and uses TFIDF for finding similar donor contexts.

In this study, we consider two different cases:
\begin{enumerate}
\item Case 1 - Context-less: we simply measure the similarity between the modification point and all repair ingredients.
\item Case 2 - Context-aware: as explained in \autoref{sec:terminology}, we consider the method as context, we measure the similarity between the recipient context surrounding the modification point, and all potential donor contexts.
\end{enumerate}

\begin{lstlisting}[language=diff,columns=flexible, frame=single, basicstyle=\footnotesize,float,label=humanpatchmath75,captionpos=b,caption={A real one statement patch, from Apache Commons Math, with high similarity between the modification point and the repair ingredient.}]
public double getPct(Object v) {            
-    return getCumPct((Comparable<?>) v);            
+    return getPct((Comparable<?>) v);            
}            
\end{lstlisting}

\subsection{Word embedding}
Word embedding is a set of techniques that map words to vectors, and with recent progress in natural language processing, the semantics of words are even preserved in the dense vector. Word2vec is one example of unsupervised word embedding technique that can be trained on a large unlabeled dataset for single word embedding \cite{mikolov2013distributed}. It is capable of generate vectors such that "king" - "man" + "woman" = "queen" \cite{mikolov2013linguistic}. The key idea is that word with similar context have a similar meaning. For example in "The quick \textbf{brown} fox jumps over the lazy dog" and "The quick \textbf{red} fox jumps over the lazy dog", the word brown and red have similar context, therefore they should also have a similar meaning. 

Doc2Vec is a word embedding techniques that generates vector representation for sequences of words like paragraphs \cite{le2014distributed}. It is built on the same idea as Word2vec, but it has a document (sequence of words) vector representation which is trained together with word embeddings. Word embedding techniques have also been successfully applied in source code at different granularities, like \cite{alon2018code2seq}, \cite{xu2017neural} and \cite{allamanis2017graph}.

\section{Research Methodology}

In this section, we describe our research methodology.
\subsection{Goals}

The goals of our methodology are the following.

\textbf{Study similarity in isolation.} While previous work \cite{ssfix-ase17,capgen-icse18,SimFix-issta2018} has used similarity analysis in the context of program repair, it was always done as part of an integrated approach with other heuristics. To our knowledge, nobody has ever studied the importance and effectiveness of similarity analysis in isolation for program repair. In this paper, we want to systematically study the impact of similarity on the search space of program repair.
    
\textbf{Study multiple similarity metrics.} All the previous work has considered a specific similarity metric well adapted to their approach. In this paper, we want to deepen the understanding of the pros and cons of different similarity metrics. Contrary to previous work, we define and systematically compare four different metrics. They are all part of the same repair process, meaning that the comparison clearly measures the impact of each of them.

\subsection{Overview of the Research Methodology}

Our research methodology is as follows.
First, we consider the search of correct repair ingredients as a ranking problem: given the modification point, return a list of repair ingredients ordered by the likelihood that they will fix the bug. In our study, the ingredient scope is the whole application, meaning that the repair ingredients are taken from the same application. 

Second, we systematically define four similarity metrics for redundancy-based program repair, which have been used by previous studies \cite{yokoyama2016toward,ssfix-ase17,white2017sorting,SimFix-issta2018}. They all capture different types of similarity.
The similarity metrics are computed between the modification point (resp. recipient context),  and all repair ingredients (resp. donor context). 

Third, we investigate the statistical difference between our similarity metrics. The rankings generated by each similarity metric are compared using the Wilcoxon signed-rank test to check they came from the same distribution, \ie if they capture the same type of similarity.

\subsection{Repair task}
In our experiments, we create ``repair tasks'' and use them for similarity analysis. A repair task is generated from a one-statement replacement commit. For each considered one-statement replacement commit, the removed statement is considered as the modification point, and the inserted statement is considered as the ground-truth correct ingredient. 
Since our focus is the role of similarity analysis in redundancy-based program repair approaches, we only keep repair tasks where the correct repair ingredient exists in the same application. In other words, all the commits in our dataset would have potential targets for redundancy-based program repair. Furthermore, we remove repair tasks where the correct repair ingredient exists in the recipient context because it is meaningless to calculate the similarity between the recipient context and the donor context when it is the same. By excluding these cases, we are focusing on the less obvious repair tasks where the repair ingredient is at least in another method, and potentially in another file.

For all repair tasks, the similarity analysis is made for repair ingredients and donor contexts from the same application, which means that the redundancy scope is the application scope. This represents a good trade-off: with smaller redundancy scope such as file scope, where repair ingredients and donor contexts are collected from the same file, the number of repair ingredients and donor context per repair tasks is too low for any meaningful similarity analysis. 

The repair tasks that we create have two differences compared to the traditional program repair workflow:
First, the protocol assumes that the fault localization step works perfectly and returns the actual modification point that caused the bug.
Second, we do not run any test. Running tests is not required because we focus on patches that are identical to the human patch. Other potential patches that are semantically equivalent to the human patch are ignored, which is a sound, conservative assumption. 
This is meant to eliminate the validity threats and the uncontrolled variables in the experiment, in order to purely focus on the effectiveness of the similarity metrics.

\subsection{Similarity Metrics for Repair}
\label{sec:similarity-metrics}

Our core idea is to compute the similarity between the modification point (recipient context) and all possible repair ingredients (donor contexts) for all repair tasks. In this paper, we consider:
\begin{enumerate}
\item Longest common subsequence (LCS)
\item Term frequency inverse document frequency (TFIDF)
\item Word embedding based on unsupervised learning (Doc2vec, \cite{le2014distributed})
\item Structure similarity between abstract syntax trees (Deckard, \cite{jiang2007deckard})
\end{enumerate} 

Our goal is that they are different in nature, and capture different characteristics of similarity. They are explained in detail in the following sections:

\subsubsection{LCS} 

LCS is purely syntactic because it works at the character level. LCS has been proposed for program repair by Yokoyama \etal \cite{yokoyama2016toward}. There exist very efficient implementations of it in all major languages. In the context of programming language, LCS is capable of capturing the similarity between words like \textit{port1} and \textit{port2}.

In our similarity analysis, LCS considers source code components as a sequence of characters, the normalized LCS is computed between pairs of source code component. 

\subsubsection{TFIDF} 

TFIDF is based on words frequency and rarity, which translates to token frequency and token rarity in the context of programs. TFIDF has been used in the context of program repair by ssFix \cite{ssfix-ase17}. Being based on tokenization, it is much less syntactic than LCS. The strength of TFIDF is its ability to focus on the unique tokens. If the variable name \textit{veryRareName} only occurred once in the file and it is in the modification point. Then, if we have an ingredient that contains \textit{veryRareName}, it is likely that ingredient is a good candidate.

In our experiments, each source code component is tokenized per the tokenization rule of the considered language (Java in our experiment). They are considered as documents and the TFIDF is calculated for each token. All source code component is then converted into vectors of TFIDF scores. Finally, the similarity score is computed by using cosine similarity, as standard practice in the field for measuring vector.

\subsubsection{Doc2vec} 

Doc2vec is a numerical vector associated with an object, meant to capture semantic relationships. In the context of programming, Doc2vec can be computed for tokens, lines, functions, \etc. DeepRepair has used a similar technique to reason about code similarities \cite{white2017sorting}. Learning an embedding such as Doc2vec on program tokens are meant to be less syntactic than LCS and TFIDF. For example, \textit{dog} and \textit{cat} should be considered semantically similar, i.e. adjacent in the vector space, while LCS and TFIDF would both consider them to be different compared to the word \textit{fog} in their respective syntactic spaces.

We train Doc2Vec on a corpus of Java files. Source code components from each java file are extracted and tokenized. The tokenized source code components are used to train a Doc2Vec model where the target vector space is 128 for repair ingredient and 300 for donor context. This results in that each source code component is associated to a vector of real numbers. Finally, the similarity score is also computed by using cosine similarity in the embedding space. 

\subsubsection{Deckard} 

Deckard generates a feature vector from each considered source code component (a statement without context, a method with context-aware similarity). The feature vector captures structural information in the considered source code component where each dimension represents a tree pattern such as the number of expression or declaration. SimFix has also used Deckard for measuring AST similarity \cite{SimFix-issta2018}.

With Deckard, we generate a feature vector for each source code component. The similarity between two feature vectors are computed by using cosine similarity.

\subsubsection{Implementation}

The following libraries was used in the experiment:
\begin{itemize}
\item JavaParser\footnote{https://javaparser.org}: To extract methods and statements from Java files.
\item javalang\footnote{https://github.com/c2nes/javalang}: To tokenize Java source code.
\item gensim \cite{rehurek_lrec}: To train Doc2vec on statements and methods.
\item Deckard \cite{jiang2007deckard}: To generate vector that represents features in AST.
\end{itemize}

\section{Experimental Methodology}

We present the design of original experiments to study the role of similarity in redundancy-based program repair.

\subsection{Research Questions}

\newcommand\rqone{How effective are the different similarity metrics on handling the search space of redundancy-based repair?}
\textbf{RQ1}: \rqone\\
In redundancy-based repair, the search space consists of all repair ingredients that can be found in a certain scope. The number of repair ingredients in that scope is sometimes overwhelming and may not be exhaustively explored. Therefore, most redundancy-based repair tools are to only select repair ingredients at random \cite{qi2014strength,weimer2009automatically}. 
The main purpose of RQ1 is to compare the four similarity metrics presented in \autoref{sec:similarity-metrics} for all repair tasks and see how they can rank the correct repair ingredient over the other.

\newcommand\rqtwo{What is the impact of considering the recipient and donor context when searching for repair ingredients?}

\newcommand\rqthree{
To what extent do the different similarity metrics rank the ingredients in the same way?}

\textbf{RQ2}: \rqtwo\\
The context around the buggy location may be useful to consider when trying to fix a bug. Several program repair approaches use the same idea when searching for the correct repair ingredient \cite{ssfix-ase17,SimFix-issta2018,capgen-icse18}: they use the recipient context to find similar donor contexts.
In this research question, we want to quantify the importance of the context by measuring the rank of the correct repair ingredient with and without context. 

\textbf{RQ3}: \rqthree\\
By having a common dataset, we can compute the statistical difference between the different similarity metrics we consider. In this research question, we look at whether two rankings of repair ingredients generated by two different similarity metrics come from the same distribution.

\subsection{Protocols}

\subsubsection{Protocol of RQ1}
\label{sec:protocol-rq1}

RQ1 is an experiment that creates repair tasks and use them to evaluate similarity analysis. Our idea is to see whether redundancy-based program repair would predict the inserted statement as the patch. This is done as follows:
\begin{enumerate}[label=\Roman*]
\item \emph{Extract repair tasks} We extract all repair tasks from a corpus from the literature, presented in Section \ref{sec:codrep} . Since the whole experiment is too computationally expensive, we take a random sample of them.
\item \emph{Collect repair ingredients}  For each modification point, all repair ingredients at application scope are collected.
\item \emph{Compute the four similarity metrics} We compute the similarity between the modification point between all ingredients. Repair ingredient that is syntactically equivalent to the modification point is excluded since they cannot be the correct repair ingredient. And if they were then similarity analysis would be meaningless since they will always be ranked first.
\item \emph{Assess similarity effectiveness} The collected scores enables us to plot the rank distribution with violin plot. The following statistical numbers are also reported:
\begin{enumerate}
\item Median rank: The median rank of the correct repair ingredient
\item Average space reduction: It is calculated by dividing the average rank of the correct repair ingredient with the average search space size of repair ingredients. With random search, we can expect that have to search half of the search space on average for finding the correct repair ingredient.
\item Perfect repair: The percentage of repair tasks where the correct repair ingredient is ranked at first.
\end{enumerate}
\end{enumerate}

\subsubsection{Protocol of RQ2}
\label{sec:protocol-rq2}

RQ2 is about quantifying the importance of context when searching for the correct repair ingredient. Our idea is to use the same repair task from \autoref{sec:protocol-rq1}, and instead of collecting all repair ingredients, we collect all donor contexts (all methods) in the application scope. The similarity is measured between the recipient context and all potential donor contexts. The procedure is:
\begin{enumerate}[label=\Roman*]
\item \emph{Collect repair tasks} All repair tasks from RQ1 are considered in order to compare the effectiveness with context and without context
\item \emph{Collect donor contexts}  For each recipient context, all donor contexts at the application scope are collected.
\item \emph{Compute the four similarity metrics} We compute the four similarity scores between the recipient context and all donor contexts. Donor contexts that are syntactically equivalent to the recipient context are excluded, since otherwise they would always be ranked first.
\item \emph{Assess similarity effectiveness} The collected scores enables us to compute the rank distribution and plot it with a violin plot. The following statistical numbers are also reported:
\begin{enumerate}
\item Median rank: The median rank of the donor context containing the correct repair ingredient
\item Average space reduction: It is calculated by dividing the average rank of the correct donor context with the average search space size of donor contexts. With random search, we can expect that have to search half of the search space on average for finding the correct donor context.
\item Perfect repair: The percentage of repair tasks where the correct donor context is ranked at first.
\end{enumerate}
\item \emph{Compare against the best context-less ranking} The ranking from RQ1 and RQ2 are not directly comparable since they are rankings of different source code components. Therefore we do the following:
\begin{enumerate*}
\item Rank all donor context with the best similarity metric found in RQ2.
\item Rank repair ingredients in each donor context with the best similarity metric found in RQ1.
\end{enumerate*}
In this way, the most similar repair ingredient in the most similar donor context will be ranked first, the second most similar repair ingredient in the most similar donor context will be ranked second and \etc. This ranking is compared against the best ranking found in RQ1, which simply ranks all repair ingredients without context.
\end{enumerate}

\subsubsection{Protocol of RQ3}
RQ3 is an experiment that computes the statistical difference between the rankings from RQ1 and RQ2. We aim to answer the question: Does different similarity metric capture different similarity? For example, we expect that Levenshtein distance and longest common distance to capture the same type of similarity, which is similarity at the character level. But in our study, we have different similarity metrics that capture syntactic, semantic and tree similarity, we want to make sure they indeed capture different similarities. 

The question is answered using the Wilcoxon signed-rank test. The absolute rank of each correct repair ingredient is used for the test. For each pair of ranking by two similarity metric, we compute the probability that the two rankings are coming from the same distribution. The test statistic and p-value from the test are reported. We choose 0.01 ($1\mathrm{e}{-2}$) as our significance level. The null hypothesis for the Wilcoxon signed-rank test is that both samples come from the same distribution. 

\subsection{Data}
\subsubsection{CodRep corpus}
\label{sec:codrep} 

In RQ1 and RQ2, we need a corpus of one statement patches for creating all repair tasks. For this, we use the CodRep corpus \cite{chen2018codrep}. The CodRep corpus contains 171605 patches from commits in 29 distinct projects in previous studies from the literature. This corpus is appropriate for our experiments because it contains 58069 unique one-line replacement patches. All one-line replacement patches are checked so that the replacement line is a statement. This corpus enables is to study redundancy-based program repair at a large-scale.

\subsubsection{Github Java Corpus}
\label{sec:github-corpus} 

Github Java Corpus is used to train statement and method embedding using Doc2vec. It is a collection of Java code at large scale, it contains 14807 Java projects from Github, which are selected as being above average quality. The above average quality is assured by filter away projects that have never been forked. All projects in GitHub Java Corpus are downloaded (clone) from GitHub and duplicate projects are removed. In total, Github Java Corpus contains 2,130,264 Java files and 352,312,696 LOC.

\subsection{Descriptive Statistics of Repair Tasks}

\begin{figure}
\begin{center}
\begin{minipage}[t]{.48\textwidth}
\includegraphics[width=\linewidth]{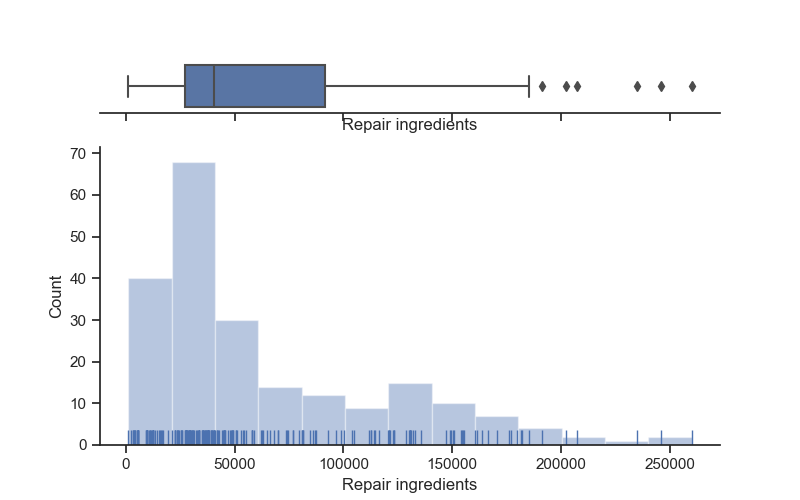}
\caption{Search space of repair ingredients for all repair tasks}
\label{fig:ingredient_dist}
\end{minipage}
\hfill
\begin{minipage}[t]{.48\textwidth}
\includegraphics[width=\linewidth]{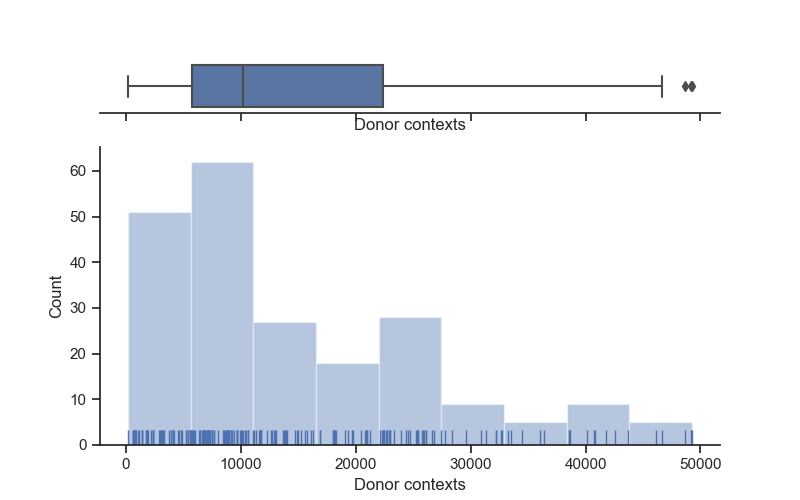}
\caption{Search space of donor contexts for all repair tasks}
\label{fig:donor_dist}
\end{minipage}
\end{center}
\end{figure}

We systematically collect program repair tasks to study the role of similarity in program repair. For each project from the CodRep corpus (presented in \autoref{sec:codrep}), we select 10 repair tasks (or less if there is not enough repair tasks satisfying our inclusion criteria).
Eventually, we have 214 repair tasks from 29 projects. 

Since the Deckard vector generation for all repair ingredient is very slow\footnote{Generating a tree representation using Deckard for say 260000 ingredients, means building 260 000 ASTs, which would take weeks and is infeasible in practice.}, therefore we only select 2 repair tasks from each project and generate vector with Deckard for all repair ingredients of those 2 repair tasks. 
For generating Deckard vector for all repair ingredients, we have in total 49 repair tasks from 29 projects. This is a strict subset of the 214 repair tasks considered.

\autoref{fig:ingredient_dist} and \autoref{fig:donor_dist} show the search space of repair ingredients and donor contexts respectively. There are on average, 63516 repair ingredients and 14452 donor contexts per repair task.
Over the 214 repair tasks, we computed \textbf{56 million similarity scores} for 15 million source code components.

\section{Experimental Results}
We now present the results of our large scale and novel experiments on the role of similarity in redundancy-based program repair.

\subsection{Research Question 1}
\emph{\rqone}

\begin{figure*}
\begin{center}
\includegraphics[width=\linewidth]{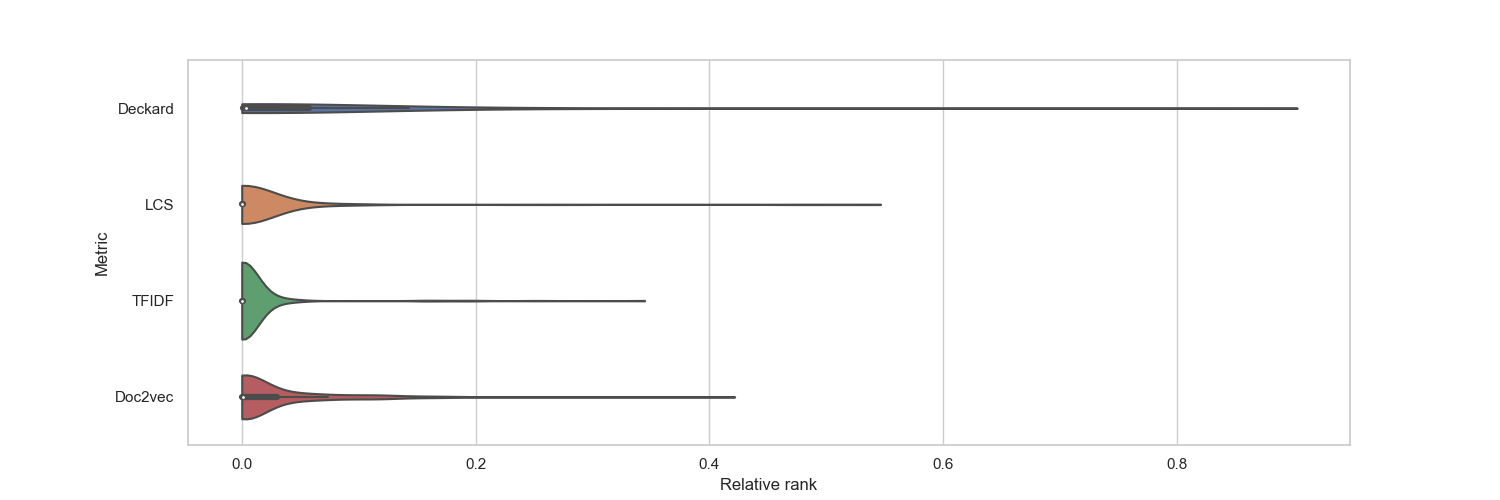}
\caption{Rank distribution of the correct repair ingredient}
\label{fig:statement_rank}
\end{center}
\end{figure*}

\begin{figure*}
\begin{center}
\includegraphics[width=\linewidth]{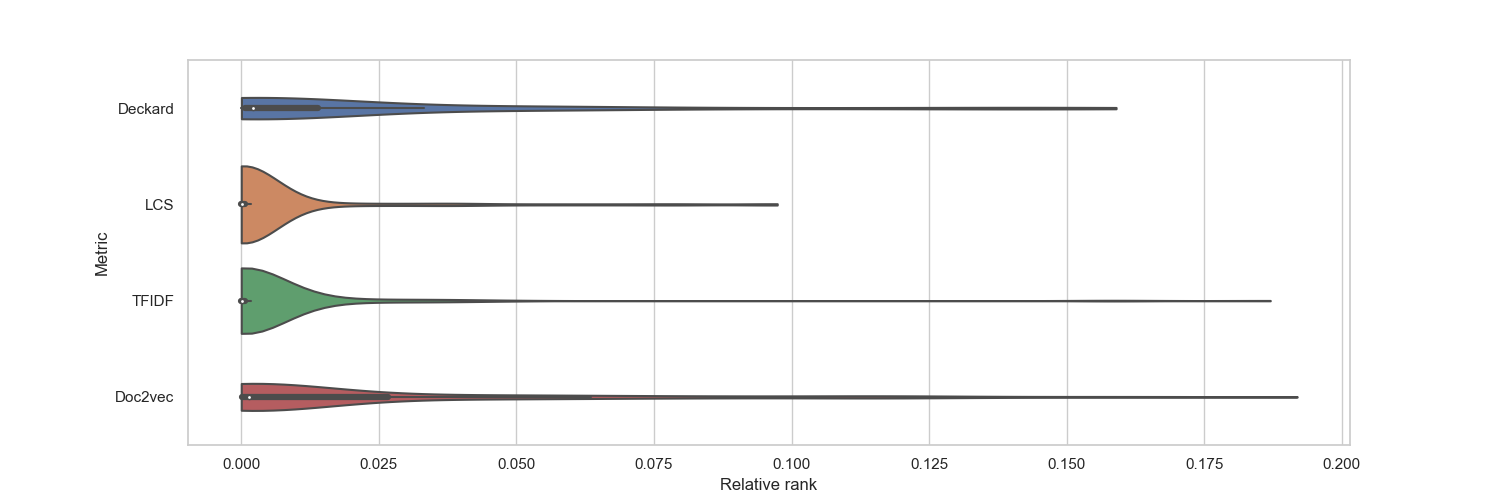}
\caption{Top 20\% rank distribution of the correct repair ingredient}
\label{fig:statement_lim_rank}
\end{center}
\end{figure*}

\begin{table}
\begin{center}
\resizebox{\linewidth}{!}{\begin{tabular}{llll}
\hline
Similarity metric & Median rank & Average space reduction & Perfect repair \\
\hline
Deckard & 165 & 91.87\% & 4\% \\
LCS & 3 & 98.17\% &  29\% \\
TFIDF & 4 & 99.34\% & 33\% \\
Doc2vec & 38 & 97.57\% & 14\% \\
Combined & 12 & 99.5\% & 17\% \\
\hline
\end{tabular}}
\caption{Rank statistic of repair ingredients}
\label{tab:statement_rank}
\end{center}
\end{table}

The rank distribution of the correct repair ingredient is shown in \autoref{fig:statement_rank} and zoomed in \autoref{fig:statement_lim_rank}. The rank is normalized by the number of repair ingredients in the program, a value of 0.1 means that the correct repair ingredient is ranked among the top 10\% of all repair ingredients. The two figures are a violin plot where they show probability density at different values. From the figures, we can see that most correct repair ingredients are ranked among the top.

\autoref{tab:statement_rank} contains the median rank in absolute numbers, the average search space reduction, along with the percentage of perfect repair.

\textbf{Rank distribution.} From \autoref{fig:statement_rank}, we can see that TFIDF has ranked the most correct repair ingredients among the top $1\%$. It has also a shorter tail than other similarity metrics, suggesting that TFIDF performed the best in general. In comparison, Deckard's performance is not good, as the rank of the correct repair ingredients are evenly distributed. One explanation for this poor performance is that the considered repair ingredients are small (one single statement) therefore the Deckard vector representing its AST  structure are too similar for most statements.

\textbf{Rank statistic.} As shown in \autoref{tab:statement_rank}, among all four similarity metrics, LCS and TFIDF outperform other metrics by a large margin in terms of percentage of perfect repairs. Recall that perfect repair means that the correct repair ingredient is ranked 1st. In those cases, the overfitting problem completely solved, since the first generated patch is the correct one. In other words, LCS and TFIDF can produce the perfect patch in 30\% of cases on average.

One possible reason why LCS and TFIDF are so effective is maybe because we consider one statement replacement patches. With small patches, string-based similarity metrics can easily capture the intended similarity, but Doc2vec and Deckard are incapable to capture it.

The reported average space reduction from \autoref{tab:statement_rank}, is calculated by dividing the mean rank of the correct repair ingredient with the average search space size of repair ingredients. Intuitively, this number means that how much repair ingredients we can avoid when using certain similarity metric. When the search space only contains one correct repair ingredient, random search would rank the correct repair ingredient in the middle of all repair ingredients (50\%) on average. But by using any of the four similarity metrics, we can at least skip 90\% of repair ingredients, the similarity metric will rank certain repair ingredient before others.

\begin{tcolorbox}
\begin{Finding}\label{find:rq1}
TFIDF performs best in general by the highest search space reduction ratio at 99.34\% and the least number of poorly performing outliers. LCS and TFIDF generated the most perfect repairs by a large margin. Deckard is not effective according to all measurements.
\end{Finding}
\begin{Implication}\label{impl:rq1}
This is directly actionable: any future implementation of redundancy-based program repair should use TFIDF as the similarity metric for finding the correct repair ingredient based on the modification point. More sophisticated approaches such as embedding and tree similarity are not effective enough.
\end{Implication}
\end{tcolorbox}

\subsection{Research Question 2}
\emph{\rqtwo}

\begin{figure*}
\begin{center}
\includegraphics[width=\linewidth]{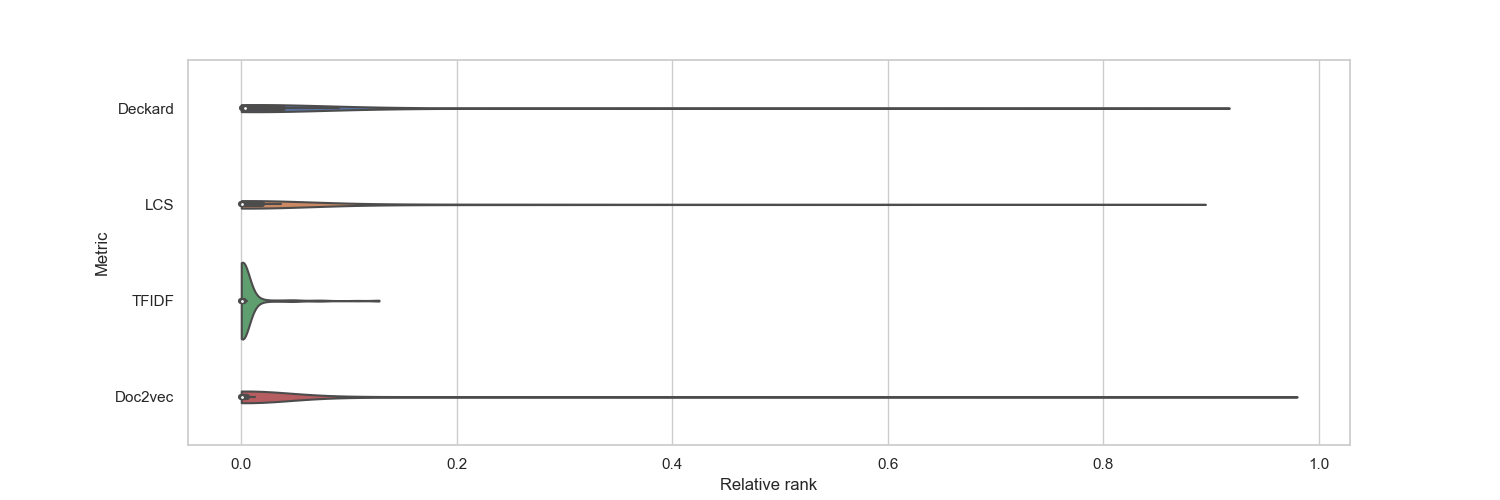}
\caption{Rank distribution of the correct donor context}
\label{fig:method_rank}
\end{center}
\end{figure*}

\begin{figure*}
\begin{center}
\includegraphics[width=\linewidth]{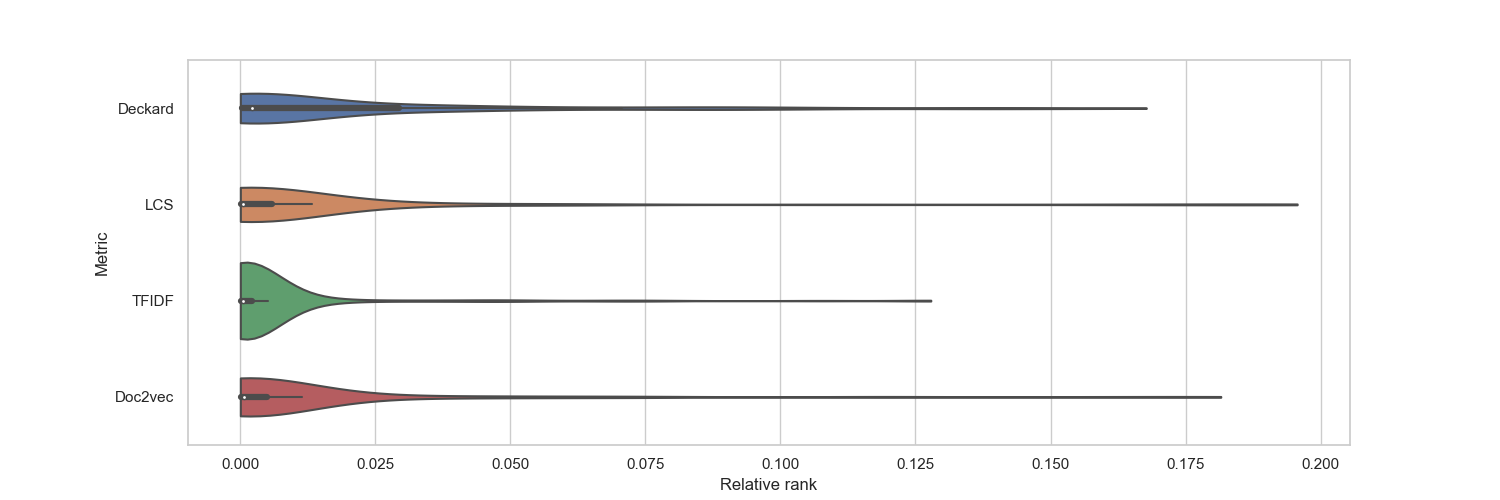}
\caption{Top 20 \% rank distribution of the correct donor context}
\label{fig:method_lim_rank}
\end{center}
\end{figure*}

\begin{table}
\begin{center}
\resizebox{\linewidth}{!}{\begin{tabular}{llll}
\hline
Similarity metric & Median rank & Average space reduction & Perfect repair \\
\hline
Deckard & 29 & 91.89\% & 13\% \\
LCS & 7 & 93.08\% & 26\% \\
TFIDF & 4 & 99.56\% & 22\% \\
Doc2vec & 6 & 96.46\% & 20\% \\
\hline
\end{tabular}}
\caption{Rank statistic of donor contexts}
\label{tab:method_rank}
\end{center}
\end{table}

The rank distribution of the correct donor context is shown in \autoref{fig:method_rank} and zoomed in \autoref{fig:method_lim_rank}. \autoref{tab:method_rank} contains the median rank in absolute numbers, the average search space reduction, along with the percentage of perfect repair.

\textbf{Rank distribution.} From \autoref{fig:method_rank}, we can see that TFIDF is the most effective in finding the right donor context, with a similar distribution and short tail compared to \autoref{fig:statement_rank}.

Interestingly, Doc2vec and Deckard achieve better results than without context. Different factors account for the increased effectiveness of Doc2Vec and Deckard: 
\begin{enumerate*}
\item Doc2vec is more effective for large documents (full method contexts) than for short document (statement), it generates more stable embedding.
\item Donor contexts contain more semantic information than repair ingredient only.
\item Donor contexts have more complex AST, meaning that the vector representing the tree can capture more information
\end{enumerate*}

On the contrary, LCS performed worse with full context than with ingredient only. It is consistent since it only focuses on syntactically similar donor context and fails to capture the semantic diversity of donor contexts. For instance, LCS cannot capture the similarity if the donor context simply reorders the tokens. TFIDF avoids this problem by ignoring the token order, and it does capture semantics by assigning different weights to each token in the donor context. 

\textbf{Rank statistic.} As shown in \autoref{tab:method_rank}, TFIDF and LCS have most perfect repairs . With context-aware similarity analysis, perfect repair means that the donor snippet containing the correct repair ingredient is ranked first. By considering donor context instead of only the repair ingredient, Deckard and Doc2vec have higher perfect repair rates (13\% versus 4\% for Deckard and 20\% versus 14\% for Doc2vec). It shows that a  bigger context is important. A bigger context means more semantically code to reason about, and it does help embedding techniques like Doc2vec. It also means that it more structural information, which benefits tree similarity metrics like Deckard.

The reported average space reduction in \autoref{tab:method_rank}, is calculated by dividing the mean rank of the correct donor context with the average search space size of donor contexts. Intuitively, this number means that how much donor contexts we can avoid when using certain similarity metric. Again, similar to the average space reduction at the repair ingredient level, all four similarity metric can at least rank the correct donor context before 90\% other donor contexts.

\begin{figure*}
    \centering
    \includegraphics[width=\linewidth]{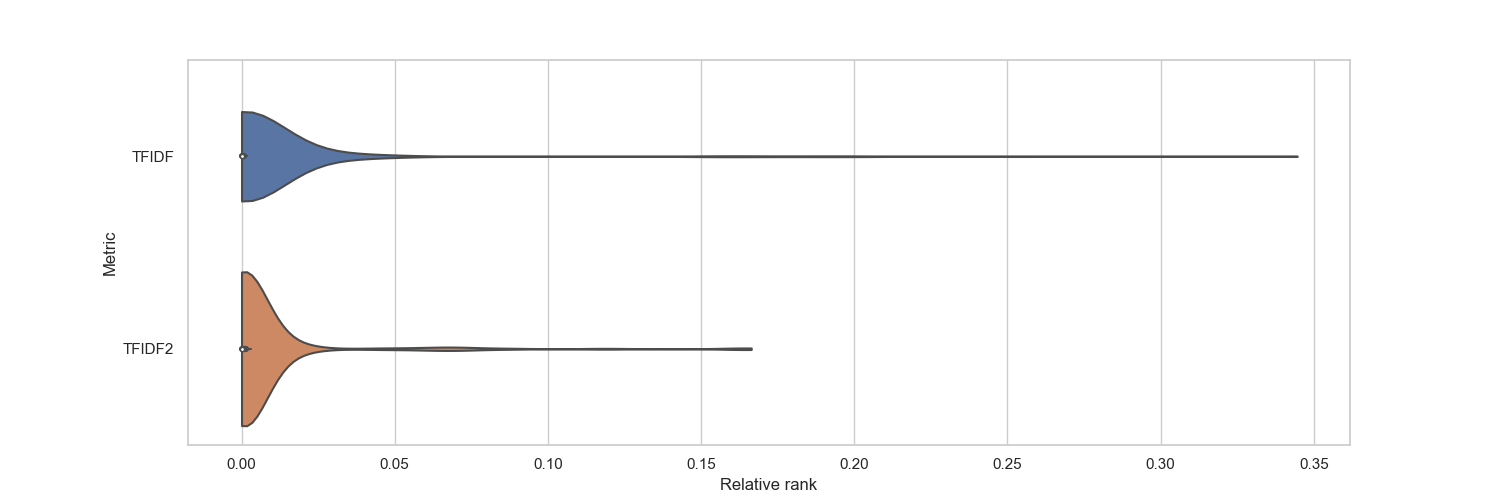}
    \caption{TFIDF ranks all repair ingredient. TFIDF2 ranks all donor contexts and then ranks repair ingredients inside each donor context.}
    \label{fig:combination}
\end{figure*}

\textbf{Combine context-less and context-aware.} The rankings from \autoref{tab:method_rank} and \autoref{tab:statement_rank} are not directly comparable, since one is a ranking of repair ingredients, and the other one is a ranking of donor contexts. We solve this issue by combining the best similarity metric for repair ingredients (TFIDF) and the best similarity metric for donor contexts (TFIDF). Meaning that the donor contexts will first be ranked by TFIDF, and the repair ingredients inside each donor context will also be ranked by TFIDF, but at the ingredient level. Then, this ranking (TFIDF2) is compared against repair ingredient ranking with TFIDF, and the result is shown in \autoref{fig:combination} and \autoref{tab:statement_rank}. 

From \autoref{fig:combination}, we can clearly see the TFIDF2 ranking improves upon simply ranking repair ingredients with TFIDF. It shows how important the context is and how it helps with ranking the correct repair ingredient. And by comparing TFIDF and TFIDF2 in \autoref{tab:statement_rank}, we can also see the TFIDF2 rank have $0.16\%$ higher space reduction rate. Considering the size of the remaining search space of TFIDF ($100 - 99.34 = 0.66$), the TFIDF2 rank cut the remaining search space by additional $0.16 / 0.66 = 24\%$. However, the median rank and perfect repair rate are lower.

\begin{tcolorbox}
\begin{Finding}\label{find:rq2}
TFIDF is the best similarity metric for identifying the best donor context. Compared to RQ1, the effectiveness of Doc2vec and Deckard improves by considering the context. By combining the best context-less ranking and the best context-aware ranking, we achieve better results than context-less ranking only.
\end{Finding}
\begin{Implication}\label{impl:rq2}
TFIDF should be used by redundancy-based program repair tools to identify donor contexts. The ranking can be further improved by rank repair ingredients inside each donor context also with TFIDF. Our results improve the external validity of previous research \cite{capgen-icse18,ssfix-ase17,SimFix-issta2018,xin2019revisiting} showing that the donor and recipient context help in identifying the correct repair ingredient.
\end{Implication}
\end{tcolorbox}

\subsection{Research Question 3}
\emph{\rqthree}

\begin{table}
\begin{center}
\begin{tabular}{lc}
\hline
Similarity metric pair & Wilcoxon signed-rank test \\
\hline
Deckard \& LCS & $n=49, T=45, p=5.99\mathrm{e}{-8}$ \\
Deckard \& TFIDF & $n=49, T=15, p=6.05\mathrm{e}{-9}$ \\
Deckard \& Doc2vec & $n=49, T=434, p=1.14\mathrm{e}{-1}$ \\
LCS \& TFIDF & $n=214,  T=6293, p=6.05\mathrm{e}{-1}$ \\
LCS \& Doc2vec & $n=214, T=5224, p=2.4\mathrm{e}{-9}$ \\
TFIDF \& Doc2vec & $n=214, T=3285, p=5.24\mathrm{e}{-15}$ \\\hline
\end{tabular}
\caption{Calculated value from Wilcoxon signed-rank test for rankings in RQ1, $n$ is the sample size, $T$ is the test statistic and $p$ is the p-value. Some cells are omitted since the table is symmetric.}
\label{tab:wilcoxon_rq1}
\end{center}
\end{table}

\begin{table}
\begin{center}
\begin{tabular}{lc}
\hline
Similarity metric pair & Wilcoxon signed-rank test \\
\hline
Deckard \& LCS & $n=211, T=6637, p=1.76\mathrm{e}{-2}$ \\
Deckard \& TFIDF & $n=211, T=2322, p=8.89\mathrm{e}{-17}$ \\
Deckard \& Doc2vec & $n=211, T=4882, p=3.37\mathrm{e}{-7}$ \\
LCS \& TFIDF & $n=214, T=3540, p=1.88\mathrm{e}{-7}$ \\
LCS \& Doc2vec & $n=214, T=6318, p=6.7\mathrm{e}{-2}$ \\
TFIDF \& Doc2vec & $n=214, T=3668, p=8.85\mathrm{e}{-6}$ \\\hline
\end{tabular}
\caption{Calculated value from Wilcoxon signed-rank test for rankings in RQ2, n is the sample size, T is the test statistic and $p$ is the p-value. Some cells are omitted since the table is symmetric.}
\label{tab:wilcoxon_rq2}
\end{center}
\end{table}

In this research question, we compute the statistical difference between the similarity metrics considered in RQ1 and RQ2. The Wilcoxon signed-rank test is used between the rankings generated by each similarity metric. The result is represented in \autoref{tab:wilcoxon_rq1} and \autoref{tab:wilcoxon_rq2}. The Wilcoxon signed-rank test returns $T$ and $p$, where $T$ is the sum of rank differences and $p$ is the p-value.

\textbf{Wilcoxon signed-rank test.}%
The results are shown in \autoref{tab:wilcoxon_rq1} for context-less similarity. We can reject null hypothesis for 4/6 similarity metrics with with significance level ($\alpha = 0.01$), which means that they capture a different property. 
For the rankings of LCS and TFIDF, the null hypothesis cannot be rejected, meaning that LCS and TFIDF capture the same type of similarity. This is reasonable considering that both metrics measure syntactic similarity.
We could also not reject the null hypothesis for rankings generated between Deckard and Doc2vec, there is not enough evidence to say that they are drawn from the same distribution.

At donor context level, the result is shown in \autoref{tab:wilcoxon_rq2}. We can reject the null hypothesis for 4/6 similarity metrics with $p<0.01$. The null hypothesis is not rejected between LCS and Deckard, as well as LCS and Doc2vec. The result is surprising since they are different similarity metrics that capture either syntactic, semantic or tree similarity. But the collected samples does not allow us the reject the hypothesis that they are coming from the same distribution when measuring donor context similarity.

\begin{tcolorbox}
\begin{Finding}\label{find:rq3}
With context-less ranking, the null hypothesis of Wilcoxon signed-rank test is rejected for all pairs of similarity metric but LCS \& TFIDF and Deckard \& Doc2vec. With context-aware ranking, the null hypothesis of Wilcoxon signed-rank test is rejected for all pairs of similarity metric but Deckard \& LCS and LCS \& Doc2vec.
\end{Finding}
\begin{Implication}\label{impl:rq3}
For those pairs of similarity metrics where we reject the null hypothesis, we can say that they capture a different type of similarity.  This suggests that future work could combine different similarity metrics to even better rank the correct repair ingredient, for instance by using learning-to-rank techniques.
\end{Implication}
\end{tcolorbox}

\subsection{Case studies}
We now present case studies of great and poor rankings done by the four similarity metrics to showcase the types of similarity that they capture. We discuss why they succeed or failed. The presented examples are taken from the experiment of RQ1.

\subsubsection{Case study 1}
\begin{lstlisting}[language=diff,columns=flexible, frame=single, basicstyle=\footnotesize,float,label=lst:aspectj13,captionpos=b,caption={An example patch from AspectJ}]
    } else if (IMessage.ERROR.isSameOrLessThan(kind)) {
setIcon(AjdeUIManager.getDefault().getIconRegistry().getErrorIcon());
    } else {
- setIcon(null);
+ setIcon(AjdeUIManager.getDefault().getIconRegistry().getInfoIcon());
    }
    if (isSelected) {
        setBackground(list.getSelectionBackground());          
\end{lstlisting}

For the patch shown in \autoref{lst:aspectj13}, the correct repair ingredient was ranked by LCS at the 13860th place out of 34407 repair ingredients, which is really bad. The modification point and the correct repair ingredient have many character level differences, therefore LCS failed to capture the similarity. However, TFIDF ranked the correct repair ingredient at the 16th place, which is much better. This shows that these two metrics have captured a different similarity for this repair task.

\subsubsection{Case study 2}
\begin{lstlisting}[language=diff,columns=flexible, frame=single, basicstyle=\footnotesize,float,label=lst:jmeter5,captionpos=b,caption={An example patch from Apache JMeter}]
        log.warn("Script did not return a value");
        return 0;
    }
-   delay = Long.valueOf(o.toString()).longValue();
+   delay = Long.parseLong(o.toString());
} catch (NumberFormatException e) {
    log.warn("Problem in JSR223 script ", e);
} catch (IOException e) {      
\end{lstlisting}

For the given patch in \autoref{lst:jmeter5}, the correct repair ingredient was ranked at the 1st place by Doc3Vec. Doc2vec was able to identify the semantic relationship between \textit{valueof().longValue()} and \textit{parseLong()}. At the same time, Deckard, TFIDF and LCS also captured the similarity relatively well by ranking the correct repair ingredient at 21st, 2nd and 2nd place. For this repair task, all similarity metrics are effective. 

\subsubsection{Case study 3}
\begin{lstlisting}[language=diff,columns=flexible, frame=single, breaklines=true, basicstyle=\footnotesize,float,label=lst:tomcat18,captionpos=b,caption={An example patch from Apache Tomcat}]
if (urlPattern == null)
    return (false);
if (urlPattern.indexOf('\n') >= 0 || urlPattern.indexOf('\r') >= 0) {
-   getLogger().warn(sm.getString("standardContext.crlfinurl",urlPattern));
+   return (false);
}
if (urlPattern.startsWith("*.")) {
    if (urlPattern.indexOf('/') < 0)
\end{lstlisting}

For the given patch in \autoref{lst:tomcat18}, all similarity metrics ranked the correct repair ingredient after 10000th place. It shows that sometimes the correct repair ingredient is very different from the modification point. It would require other kinds of program analysis to find the correct repair ingredient, such as mutation operation prioritization \cite{capgen-icse18}.

\section{Related work}

\subsection{Redundancy in Programs}
After the initial successes of GenProg, studies have looked at the underlying redundancy assumption. Barr \etal checked it against 12 Apache project, and found that changes are $43\%$ redundant at the line level \cite{barr2014plastic}.
Martinez \etal measured redundancy the at line and token level for 6 projects. They found that $3-17\%$ of commits are redundant at the line level \cite{Martinez:2014:FIA:2591062.2591114}.
Sumi \etal further conducted redundancy experiments using a larger dataset and obtain similar results  \cite{sumi2015toward}. Lin \etal examined code redundancy for 2640 Java projects with different token lengths for several types of code constructs, they studied how it affects the performance of code completion \cite{lin2017uniqueness}.
Gabel and Su investigated the opposite property, which is the uniqueness of source code \cite{gabel2010study}. They found that software lacks uniqueness at the granularity of one to seven lines of code.

Repair approaches based on code exploits redundancy in a more functional way. Xiong \etal use code search on Github to find snippets \cite{xiong2017precise} .
The approach by Ke \etal search for existing code snippets (i.e. redundant ones that match a given input-output specification \cite{KeASE2015}.

\subsection{Analyses of the Repair Search Space}

Martinez and Monperrus analyzed the repair actions over commits for 14 Java projects \cite{Martinez2013}. They showed that certain repair actions are more common than others, \textit{statement insertion of method invocation} is for instance the most common repair action. They analyze the search space of program repair with respect to those repair actions.
Our study is different compared to theirs because we concentrate on repair ingredients and donor contexts while they focus on combinations of repair actions (what they call repair shapes).

Qi \etal studied how random search compares to genetic programming to guide program repair through the search space \cite{qi2014strength}. They showed that in most cases, random search outperforms GenProg. While they only focus on the first plausible patch, we only consider the human patch, and we showed that similarity analysis is much better than random search.

Long and Rinard analyzed the search space for several repair systems \cite{long2016analysis}. They found that correct patches are sparse in the search space, while plausible patches are abundant. They also showed that increasing the size of search space decreases the ability to find the correct patch.
Our study has provided a solution to handle larger search space by using similarity analysis to reduce the search space. We have also shown that similarity analysis and rank the correct repair ingredient first, completely avoid all other plausible patches in the search space.

\subsection{Similarity in Program Repair}

Ji \etal is possibly the first to have proposed that the repair ingredients should be taken from similar code \cite{Ji2016AutomatedPR}.
Xin and Reiss further built on this idea and proposed TFIDF to compute two similarities \cite{ssfix-ase17}:
the similarity between the ingredients and their respective context, and the buggy statement and its context. In the following paper \cite{xin2019revisiting}, they have changed how they define the context and similarity based on empirical study and achieved much better result.
White \etal uses deep learning to reason about the similarity between the method body containing the modification point and the method body of ingredients \cite{white2017sorting}. 
Tanikado \etal proposed the original idea of looking at how fresh repair ingredients are, where freshness is defined by on the last updated time \cite{tanikado2017new}. Compared to our study, the difference is that they consider a constant-sized region for each program statement while we consider methods.
Wen \etal prioritized ingredients that have similar programming context based on program analysis \cite{capgen-icse18}. 
Jiang \etal considers three different similarity levels, structure similarity, variable name similarity and method name similarity \cite{SimFix-issta2018}. The final similarity score is the sum of the three similarities.
Those works have innovatively used similarity for program repair and we build on their initial results. Yet, they all focus on proposing a new repair technique and do not systematically analyze the search space. On the contrary, our paper is a principled study dedicated to the role of similarity in redundancy-based program repair.

\section{Conclusion}
We have performed an original study of the search space of redundancy-based program repair and in particular, the role of using similarity analysis to explore it. Our findings show that similarity can effectively reduce the search space in order to find the correct repair ingredient. And incorporating context to similarity analysis has a different impact depending on the similarity metric. In general, TFIDF performed the best where it could rank the correct repair ingredient and donor context in the top 1\%. And it is able to rank the correct repair ingredient and donor context first in 30\% of cases, which means that it contributes to avoiding overfitting ingredients in the search space.

\printbibliography

\end{document}